\documentclass[aps,pra,superscriptaddress,twocolumn,tightenlines,showpacs,floatfix,amsmath,amssymb]{revtex4-1}

\usepackage{physics}
\usepackage{mathtools}
\usepackage{graphicx}
\usepackage{dcolumn}
\usepackage{bm}
\usepackage{bbold}
\usepackage{xcolor}
\usepackage{hyperref}

\begin{document}

\title{Quantum state reduction of general initial states through spontaneous unitarity violation}

\author{Aritro Mukherjee}
\affiliation{Institute for Theoretical Physics Amsterdam,
University of Amsterdam, Science Park 904, 1098 XH Amsterdam, The Netherlands}
\author{Srinivas Gotur}
\affiliation{Institute for Theoretical Physics Amsterdam,
University of Amsterdam, Science Park 904, 1098 XH Amsterdam, The Netherlands}
\author{Jelle Aalberts}
\affiliation{Institute for Theoretical Physics Amsterdam,
University of Amsterdam, Science Park 904, 1098 XH Amsterdam, The Netherlands}
\author{Rosa van den Ende}
\affiliation{Institute for Theoretical Physics Amsterdam,
University of Amsterdam, Science Park 904, 1098 XH Amsterdam, The Netherlands}
\affiliation{Universit\'e Paris 1 Panth\'eon-Sorbonne,
2 Rue Cujas, 75231 Paris cedex 05, France}
\author{Lotte Mertens}
\affiliation{Institute for Theoretical Physics Amsterdam,
University of Amsterdam, Science Park 904, 1098 XH Amsterdam, The Netherlands}
\affiliation{Institute for Theoretical Solid State Physics, IFW Dresden, Helmholtzstr. 20, 01069 Dresden, Germany}
\author{Jasper van Wezel}
\affiliation{Institute for Theoretical Physics Amsterdam,
University of Amsterdam, Science Park 904, 1098 XH Amsterdam, The Netherlands}

\date{\today}

\begin{abstract}
The inability of Schr\"odinger's unitary time evolution to describe measurement of a quantum state remains a central foundational problem. It was recently suggested that the unitarity of Schr\"odinger dynamics can be spontaneously broken, resulting in measurement as an emergent phenomenon in the thermodynamic limit. Here, we introduce a family of models for spontaneous unitarity violation that apply to generic initial superpositions over arbitrarily many states, using either single or multiple state-independent stochastic components. Crucially, we show that Born's probability rule emerges spontaneously in all cases.
\end{abstract}

\maketitle


\section{Introduction\label{sec:0}}
How the unitary time evolution prescribed by Schr\"odinger's equation can be reconciled with the observation of single measurement outcomes randomly selected according to Born's probability distribution, remains one of the central foundational problems of modern science~\cite{Bassi_03_PhyRep, leggett2005quantum, overview, arndt2014testing, carlesso2022present}. One way to formulate this `quantum measurement problem', is to observe that one registers a single outcome upon performing a single quantum measurement. Repeating the measurement with the same initial state might yield a different outcome, in accordance with Born’s rule~\cite{Born26}. Describing the measurement device as a macroscopic collection of interacting quantum particles, however, its evolution should be governed by Schr\"odinger's equation. As formalized by Von Neumann~\cite{Von_Neumann2018-bo}, the interaction between a measurement device $\ket{M}$ and microscopic quantum system $\ket{S}$ in the so-called strong measurement limit, then inevitably leads to the prediction of an entangled state between system and measurement device of the form:
\begin{align}
     \Big(\sum_j \alpha_j \ket{S_j} \Big) \ket{M} \to  \sum_j \alpha_j \ket{S_j}\ket{M_j}
     \label{eq:init_superposition}
\end{align}
Although ever more massive objects have successfully been put into spatial superposition~\cite{Arndt_Nat_99, Hackermuller_PRL_2003, Gerlich_NatComm_2011, Gasbarri_ComminPhys_21}, there is no evidence of truly macroscopic measurement machines ending up in the superposition of measurement outcomes described by Eq.~\eqref{eq:init_superposition} during individual experiments.

Attempts to theoretically address the measurement problem can be grouped into three broad categories. The first posits that decoherence may be seen as a type of measurement, because it leads to diagonal reduced density matrices after tracing out the environment~\cite{Zurek_1982, Schlosshauer_2005,Zurek2009,theo13}. This approach, however, is explicitly restricted to describing expectation values averaged over an ensemble of realisations of the environment, and hence does not resolve the issue of a single outcome being observed in a single measurement~\cite{Bassi_03_PhyRep,Dieks_1989,Adler_2003,Stillfried_2008,Fortin2014}.

Second are the interpretations of quantum mechanics, which all share the central assumption that Schr\"odinger’s equation (and hence unitary dynamics) applies without change to all objects in the universe, large or small~\cite{Everett_1957,Bohm52A,Bohm52B, rovelli1996relational,Fuchs_2014}. These theories then give different interpretations for the physical meaning of the quantum state to explain why the superposed states of macroscopic objects that are unavoidable under unitary dynamics are not observed in our everyday experience. Since all interpretations strictly adhere to Schr\"odinger’s equation, the predictions from different interpretations for any given experiment are all identical, and they cannot be experimentally distinguished or verified. Notice however, that any experimental observation of Schr\"odinger's equation being violated would suffice to falsify all interpretations.

In contrast, the third class of approaches, which introduce objective collapse or dynamical quantum state reduction (DQSR) theories, share the common assumption that the quantum state does represent the actual state of physical objects of any size, and that the observed emergence of classical physics necessitates a refinement of Schr\"odinger's equation~\cite{BohmBub_66_RevModPhys, Pearle_76, Gisin84, Ghirardi_1986, Diosi_87_PLA,Ghirardi_90_PRA,percival95, Penrose_96,Wezel10, Snoke2021,Snoke2023,mukherjee2023colored}. These theories introduce small modifications to quantum dynamics that have no noticeable effect on the microscopic scale of elementary particles, but which begin to influence the dynamics in a mesoscopic regime (defined differently in different theories, but roughly understood to involve objects of beyond $10^6$ atoms being superposed over distances comparable to their own size~\cite{Penrose_96}). Beyond the quantum-classical crossover, in the macroscopic world of human measures, the result is a nearly instantaneous, dynamical reduction of the quantum state to a single, classical configuration. Because these theories introduce actual changes to the laws of quantum dynamics at the mesoscopic level, they provide experimentally testable predictions, which are a target of active and ongoing investigation~\cite{overview, Bouwmeester_2003, underground, vinante_mezzena_falferi_carlesso_bassi_2017, carlesso_bassi_falferi_vinante_2016,Snoke2023Expt}.

In this article, we generalize the recently suggested idea that spontaneously broken unitarity can cause quantum measurement~\cite{Wezel10, Mertens_PRA_21, Mertens22}, and we show that it gives rise to a family of objective collapse theories describing the measurement of generic initial states. These models differ from existing objective collapse theories in two essential ways. First, the modified quantum state evolution is
continuous and (once) differentiable, in contrast to the evolution encountered in other theories~\cite{Bassi_03_PhyRep}, which is either non-differentiable (but continuous), such as in the Di\'{o}si-Penrose or Continuous Spontaneous Localization (CSL) models~\cite{Diosi_87_PLA,Ghirardi_90_PRA,percival95,Pearle_76,Gisin84}, or contains discontinuous stochastic jumps such as in the Ghirardi-Rimini-Weber (GRW) model~\cite{Ghirardi_1986}. Secondly, although any collapse evolution necessarily involves both a non-linear and a stochastic component~\cite{Mertens_PRA_21}, these are strictly separated in the models introduced here, and the distribution of the stochastic term is independent of the state being measured. This ensures that Born’s rule emerges spontaneously in the thermodynamic limit without being assumed in the proposed modifications to quantum dynamics~\cite{Mertens22}. For a more extensive summary of the general theory of spontaneous unitarity violation, and its relation to spontaneous symmetry breaking, see Appendix~\ref{app:SSB}.

In Sec.~\ref{sec:1}, we briefly review how Spontaneous Unitarity Violations (SUV) lead to DQSR in the ideal measurement setup starting from a two-state superposition. In \ref{Sec:Ns1RV}, \ref{Sec: N state N-1 RV} and \ref{Sec:NslogNRV} we generalize this initial result and explicitly construct DQSR models for generic initial states consisting of $N$-component superpositions. We discuss three ways of introducing the required stochastic component into the $N$-state dynamics, leading to models with either a single, $N$, or $\log(N)$ random variables. We conclude in Sec.~\ref{Sec:conclusions} with a brief comparison and discussion of these models for quantum state reduction resulting from spontaneous unitarity violation.


\begin{figure}
\includegraphics[width=\columnwidth]{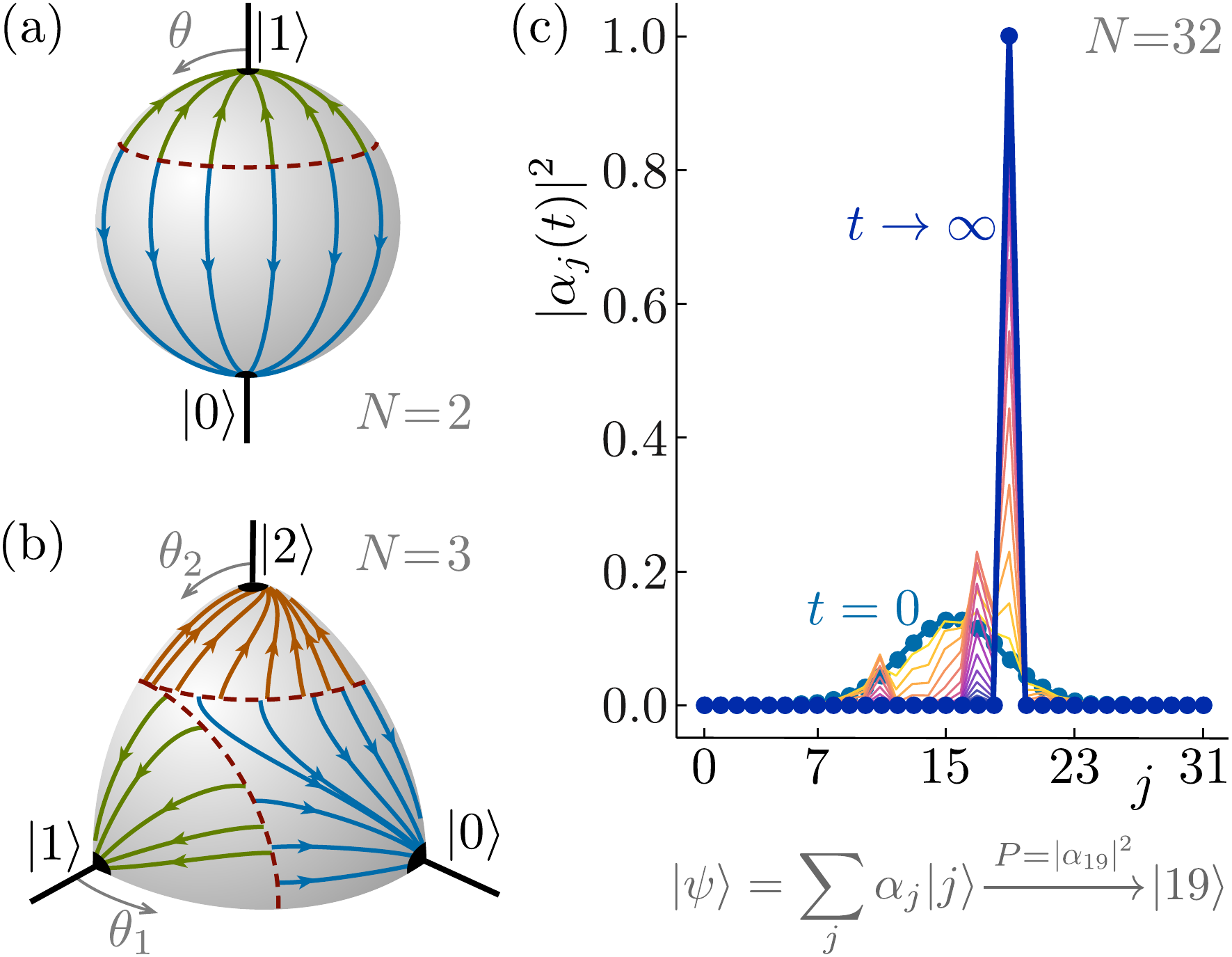}
\caption{Dynamics of quantum state reduction. (a) The state evolution of superpositions of two pointer states as given by Eq.~\eqref{eq:thetadot}, depicted on the Bloch sphere. The pointer states form attractive fixed points of the flow on the poles of the Bloch sphere. The position of the dashed red separatrix is determined by the value of the stochastic variable $\xi$. (b) Generalization of the evolution to superpositions of three pointer states (extreme points in the flow), as given by Eq.~\eqref{Eq:thetadotN}. (c) Example of an initial state superposed over eight pointer states $\ket{j}$, being dynamically reduced (for a single value of the stochastic variable) to the final measurement outcome $\ket{2}$. The probability that the randomly chosen stochastic variable leads to this particular outcome is given by $P=|\alpha_2|^2$, in accordance with Born's rule.}
\label{fig1}
\end{figure}


\section{\label{sec:1} Quantum state reduction from spontaneous unitarity violations}
In this section, we briefly review the application of spontaneous unitarity violation to the quantum measurement problem~\cite{Wezel10, Mertens_PRA_21}. Following Von Neumann~\cite{Von_Neumann2018-bo}, we consider a strong measurement setup in which a microscopic system and macroscopic apparatus are instantaneously coupled and brought into the entangled state of Eq.~\eqref{eq:init_superposition}. (see appendix~\ref{app:vonNeumann} for a more detailed description of this process). From here on, we will consider the joint evolution of the system and measurement device, and label their combined states $\ket{\psi_i}\equiv \ket{S_i}\ket{M_i}$, representing both the microscopic system being in state $\ket{S_i}$ and the measurement apparatus being in state $\ket{M_i}$. Notice that the states of the measurement apparatus in this expression are not arbitrary. As described below, the effect of the spontaneous unitarity violation will be to always reduce macroscopic objects to states with a spontaneously broken global symmetry, or equivalently, an order parameter~\cite{Wezel10} Only those macroscopic systems that are already susceptible to ordering, turn out to be susceptible to spontaneous unitarity violations. This implies that only objects normally referred to as ``classical objects", such as tables, chairs, pointers, magnets, and superconductors~\cite{Wezel_SSBlecturenotes}, act as measurement devices. It also implies that the states $\ket{M_i}$ must be states with a spontaneously broken symmetry. These include states with a well-defined position such as an actual pointer, which breaks translational symmetry. On the other hand, they exclude states with a well-defined total momentum, which cannot be stabilised in any natural process of spontaneous symmetry breaking~\cite{Wezel_SSBlecturenotes}. The observation that only ordered states are susceptible to unitarity breaking perturbations (even if all states may be subjected to such perturbations) thus imposes a preferred basis for the states of the measurement apparatus. Incidentally, states with broken global symmetry, such as actual pointers, are stable under interactions with the environment, and would be classified as ``pointer states" in the language of the theory of decoherence~\cite{Zurek_1981}. Since they represent states of classical pointers both in the sense of symmetry breaking, and in that of decoherence, we will refer to the states $\ket{M_i}$ simply as pointer states from here on. An evolution starting from the superposition of pointer states in Eq.~\eqref{eq:init_superposition}, and ending in a single state $\ket{\psi_i}$, then constitutes a description of quantum measurement.

\paragraph*{\textbf{Requirements}}
~Any theory of DQSR necessarily includes a stochastic element in order to allow for the same initial state to yield different measurement outcomes in repeated experiments~\cite{Bassi_03_PhyRep,overview}. Furthermore, because the probability of finding any particular measurement outcome depends on the initial state, the DQSR dynamics must also necessarily be a state-dependent and thus non-linear process~\cite{Mertens_PRA_21}. Finally, in order to obtain irreversible single-state dynamics and stable end points of the quantum measurement process, it must be non-unitary~\cite{Wezel10, Mertens_PRA_21}.

A non-unitary measurement process necessarily implies the breakdown of time inversion symmetry, in the sense that the probabilistic prediction of measurement outcomes based on the initial state differs from the assignment of initial state likelihoods based on a given measurement outcome (notice the difference with time reversal symmetry: a magnet in equilibrium spontaneously breaks time reversal symmetry. The magnetized equilibrium configuration, however, is static and thus evolves the same way under time evolution forwards and backwards in time. That is, its dynamics still has time inversion symmetry). The central idea of introducing spontaneous unitarity violations (SUV), is that time inversion symmetry can be broken spontaneously, in the same way that any other symmetry of nature can be spontaneously broken. That this is possible, is signalled by the diverging susceptibility of Schr\"odinger dynamics to infinitesimal non-unitary perturbations in the thermodynamic limit~\cite{Wezel_2008}. As usual in descriptions of spontaneous symmetry breaking (see appendix~\ref{app:SSB} for details), this signals a separation between the behaviour of microscopic and macroscopic objects. Single, microscopic quantum particles will not be noticeably affected within the age of the universe by the presence of a small unitarity-breaking perturbation to Schr\"odinger's equation. On the other hand, rigid macroscopic objects, which consist of a macroscopic number of quantum particles that together break a global symmetry, the effect of even the weakest unitarity-breaking perturbation is large and nearly instantaneous. 

The singular limit describing the dichotomy between the time evolution of microscopic and macroscopic objects is typical of spontaneous symmetry breaking, and emergence in general. Notice that in contrast to what the name suggests, the breakdown of unitarity, and symmetry in general, is not actually ``spontaneous"~\cite{Wezel_SSBlecturenotes}. Any large but finite-sized object requires a small but non-zero perturbation to break a symmetry. The process is called spontaneous, because for objects on human scales, the number of quantum particles collectively forming classical objects is so large, that we can never hope to detect or control the unimaginably weak perturbations that suffice to break their symmetries. As long as symmetry-breaking perturbations are not forbidden by any physical law, they will be present in some nearly-infinitesimal amount and have a large and unavoidable effect on macroscopic objects~\cite{Wezel_SSBlecturenotes}. 

In the case of SUV, it is known that unitarity is not a fundamental property of our universe, as testified for example by general relativity not being invariant under time inversion symmetry and not allowing for a description in terms of unitary time evolution~\cite{Penrose_96}. The diverging susceptibility to non-unitary perturbations therefore unavoidably causes sufficiently macroscopic objects to violate the unitarity of Schr\"odinger dynamics and be reduced to classical, symmetry-breaking states~\cite{WezelBerry}. The time scale over which the quantum state reduction takes place scales inversely with the size of the order parameter, and can thus be immeasurably small for macroscopic, ordered objects while remaining longer than the age of the universe for microscopic or non-rigid objects without an order parameter. In between these limits, there must then exist a regime of mesoscopic objects that evolve non-unitarily over human time scales.

Finally, adding a stochastic component to the non-unitary perturbation yields an objective collapse model for quantum measurement, starting from initial state superpositions of the form of Eq.~\eqref{eq:init_superposition} and evolving to different classical measurement outcomes with different probabilities. In this article, we study the long-time statistics of the classical states realised in such stochastic models for spontaneous unitarity violation, rather than studying their microscopic origin or making quantitative predictions for the time evolution during measurement. We show that Born's rule can spontaneously emerge from the stochastic dynamics, in the sense that it arises from a process driven by random variables whose distribution is independent of the quantum state being measured.

\paragraph*{\textbf{Modified Schr\"odinger equation}}
~To be specific, consider the time evolution generated by the modified Schr\"odinger equation:
\begin{align}
i\hbar \frac{\partial|\psi(t)\rangle}{\partial t} =  [\hat{H} + i\epsilon \mathcal{N} \hat{G}]|\psi(t)\rangle.
\label{Eq:key}
\end{align}
Here $\hat{H}$ is the standard Hamiltonian acting on the joint state $\ket{\psi}$ of the microscopic system and measurement device. The unitarity-breaking perturbation is written as $\epsilon \mathcal{N} \hat{G}$, making explicit that it couples to an order parameter of the measurement device and hence scales extensively with its size $\mathcal{N}$~\cite{Wezel_2008}. Moreover, its strength $\epsilon$ is taken to be nearly infinitesimal, so that it has negligible effect on the dynamics of microscopic systems while affecting an almost instantaneous evolution in the limit of large system size. The operator $\hat{G} := \hat{G}({\psi(t), \xi(t)})$ is Hermitian but non-linear and depends on the state $|\psi(t)\rangle$ as well as the instantaneous value of a time-dependent stochastic variable $\xi(t)$. Together with a specification of the dynamics for $\xi(t)$, Eq.~\eqref{Eq:key} describes a Markovian quantum state evolution. Notice, however, that this non-unitary dynamics describes the full state of the joint system and is not an effective model. It differs in this respect from the standard Gorini-Kossakowski-Sudarshan-Lindblad (GKSL) master equations, obtained for example by tracing out an environment in open quantum systems~\cite{Lindblad1976,GKS76}.

In contrast to many other models for DQSR, we do not assume the stochastic variable $\xi(t)$ to be Gaussian white noise, and $\xi(t)dt$ is not the infinitesimal Wiener measure $dW_t$~\cite{Bassi_03_PhyRep}. Instead, we assume that the stochastic variable has a non-zero correlation time $\tau$, and we will be mostly interested in the thermodynamic limit $\mathcal{N}\to\infty$, in which the state $\ket{\psi(t)}$ evolves much faster than the stochastic variable. In that limit $\tau$ is effectively infinite and $\xi(t)$ can be taken to be a time-independent variable that is randomly chosen from a stationary distribution for each realisation of the quantum measurement process. 

\paragraph*{\textbf{Two-state superpositions}}
Specialising to initial states superposed over pointer states, as in Eq.~\eqref{eq:init_superposition}, we can take the Hermitian part $\hat{H}$ to be zero, because all pointer states of a good measurement device should become degenerate eigenstates of the Hamiltonian in the thermodynamic limit~\cite{Wezel_SSBlecturenotes}. Furthermore, the non-unitary contribution to the dynamics, $\hat{G}$, must couple to the order parameter describing the broken symmetry of the pointer state in order for the process of spontaneous unitarity violation to take effect~\cite{Wezel_2008,Wezel10}. It must thus be diagonal in the pointer state basis and have different eigenvalues for different pointer states. The minimal way in which all requirements on $\hat{G}$ can be implemented for the specific case of a two-state superposition, is to consider:
\begin{align}
\ket{\psi(t)} &= \alpha(t)\ket{0}+\beta(t)\ket{1} \notag \\
\hat{G} \ket{\psi(t)} &= \bigg(\langle\hat{\sigma}_z\rangle\,+\xi\,\bigg)\hat{\sigma}_z \ket{\psi(t)}
\label{2 state eq1}
\end{align}
In this expression, $\hat{\sigma}_z:=\ket{0}\bra{0}-\ket{1}\bra{1}$ and $\langle \hat{\sigma}_z\rangle = \bra{\psi}\hat{\sigma}_z\ket{\psi}/\braket{\psi}{\psi} = \frac{|\alpha(t)|^2-|\beta(t)|^2}{|\alpha(t)|^2+|\beta(t)|^2}$, which is the usual time-dependent quantum expectation value. The coupling to the order parameter ($\langle \hat{\sigma}_z\rangle$) appears in a non-linear way (depends on the wave-function), allowing the pointer states to be stable end states of the non-unitary evolution~\cite{Mertens_PRA_21}. The stochastic variable $\xi$ is taken from a flat, uniform distribution on the interval $[-x,x]$, with $x$ a parameter whose value will be determined below. Notice that $\xi(t)$ evolves independently from $\ket{\psi(t)}$, and represents a separate physical process that is not influenced in any way by the quantum state evolution. That is, the combination of the stochastic term in Eq.~\eqref{2 state eq1} being linear and its probability density function not depending on $\ket{\psi}$ ensures that Born's rule is not imposed in the definition of the stochastic evolution and instead has to emerge spontaneously~\cite{Mertens22}. This is contrary to other models for DQSR, in which the stochastic term is multiplied by an expectation value, and thus obtains a state-dependent probability distribution that enforces Born's rule~\cite{Bassi_03_PhyRep}.

The time evolution implied by Eqs.~\eqref{Eq:key} and~\eqref{2 state eq1} does not conserve the norm of $\ket{\psi}$. This is not a problem as all physically observable expectation values can be defined in a norm-independent way as $\langle \hat{O}\rangle = \bra{\psi}\hat{O}\ket{\psi}/\braket{\psi}{\psi}$~\cite{Mertens_PRA_21}. Alternatively, and equivalently, the time evolution can be augmented with a normalisation of the wave function either at each time step $dt$ or at the end of a period of evolution, as in other models for DQSR~\cite{Bassi_03_PhyRep}. To be explicit, a normalization prescription may be obtained by noting that in the limit of interest, where the quantum state dynamics is much faster than the noise dynamics, we may consider $\xi$ to be time independent random number, sampled once in each measurement. In this limit, there is no distinction between the It\^o and Stratonovich implementations of stochastic evolution~\cite{Ito_strat1,Hanggi94_ito_strat2,mukherjee2023colored}, and the usual rules of calculus apply. Thus, the time evolution can be made norm preserving by 
adding a normalising factor to the time evolution operator.
Written in terms of the generator $\hat{G}$, this implies adding a (non-linear) term proportional to the identity operator, leading to the explicitly norm-preserving expression:
\begin{align}
\hat{G} \ket{\psi(t)} = \bigg(\langle\hat{\sigma}_z\rangle\,+\xi\,\bigg)\bigg[\hat{\sigma}_z-\langle\hat{\sigma}_z\rangle\bigg] \ket{\psi(t)}
\label{2 state eq2}
\end{align}
Notice that Eqs.~\eqref{2 state eq1} and~\eqref{2 state eq2} yield precisely the same predictions for all physically observable expectation values $\langle \hat{O}\rangle = \bra{\psi}\hat{O}\ket{\psi}/\braket{\psi}{\psi}$. For situations in which $\xi$ is time-dependent on the scale of the quantum state evolution, technical details regarding the so called quadratic-variation of the quantum state dynamics must be taken into account in order to obtain norm-preserving dynamics. A treatment of this general case may be found in Ref.~\cite{mukherjee2023colored}, but is not required in the present discussion.

Notice that the dynamics given by Eq.~\ref{2 state eq2} is distinct from the so-called continuous spontaneous localization (CSL) models and other related models driven by white noise~\cite{Diosi_87_PLA,Ghirardi_90_PRA,percival95,Pearle_76,Gisin84,Bassi_03_PhyRep}. Furthermore, it is also distinct from the spontaneous collapse models proposed in Refs.~\cite{Snoke2021,Snoke2023,Snoke2023Expt}, which have only stochastic terms while the dynamics in Eqs.~\eqref{2 state eq1} and~\eqref{2 state eq2} crucially depends on both a stochastic term and a purely deterministic non-linear term. 

To generalize Eq.~\eqref{2 state eq2}, the issues of having to define the unobservable norm and total phase of $\ket{\psi(t)}$ can be circumvented by focusing on only the physical content of the state $\ket{\psi}$, represented by the Euler angles $\theta$ and $\varphi$ defining its representation on the Bloch sphere (see Fig~\ref{fig1}). In fact, the relative phase $\varphi$ does not influence the evolution of $\theta$ for the time evolution generated by Eq.~\eqref{2 state eq1} and Eq.~\eqref{2 state eq2}. We thus restrict attention to only the dynamics of the relative weights, given by~\cite{Mertens_PRA_21}:
\begin{align}
    \label{eq:thetadot}
    \hbar\, {d{\theta}}/{dt}=\epsilon \mathcal{N}\sin(\theta) \left( \xi - \cos(\theta) \right)
\end{align}
Notice that the change in $\theta$ from time $t$ to $t+dt$ is completely specified by the values of $\theta$ and $\xi$ at time $t$ itself. The time evolution is thus a Markovian process without memory~\cite{Bassi_03_PhyRep}. Moreover, because the value of the stochastic variable $\xi$ is newly sampled for every realisation of the measurement process, the time evolution cannot be used for quantum state cloning, despite being non-linear~\cite{Wootters_Nat_82,DIEKS1982}.

The non-linear dynamics on the Bloch sphere defined by Eq.~\eqref{eq:thetadot} has stable fixed points at $\theta = 0$ and $\theta=\pi$, which represent the two pointer states appearing in the initial state superposition. It also has an unstable fixed line separating the attractive fixed points (a separatrix) at $\theta =\cos^{-1}(\xi)$, as shown in Fig.~\ref{fig1}. If the value of the randomly sampled variable $\xi$ is such that the initial value $\theta(t=0)\equiv\theta_0$ lies above the separatrix, the state evolves towards $\theta=\pi$ under the non-unitary time evolution, while it evolves towards $\theta=0$ otherwise. The probability for ending up at either pole is thus determined by the probability for the randomly selected value $\xi$ to be smaller or larger than $\cos(\theta_0)$. Choosing the range from which $\xi$ is sampled to be $[-1,1]$ results in final state statistics equaling Born's rule~\cite{Mertens_PRA_21, Mertens22}. This ensures the emergence of Born's rules in Eq.~\eqref{eq:thetadot} and Eq.~\eqref{2 state eq2} for uniformly distributed $\xi$ and this property will be utilized to construct more general models in the following sections. Notice that restricting $\xi$ to be sampled from a bounded domain restricts the type of underlying physical processes that may give rise to the stochastic evolution $\xi(t)$. It does not, however, introduce a state-dependence in the value or probability distribution of $\xi(t)$, and thus does not impose Born's rule in the definition of the stochastic variable.

With the choice $x=1$, the time evolution of Eq.~\eqref{2 state eq1} defines a model for DQSR starting from a two-state superposition in the initial state. The spontaneous breakdown of unitarity takes place in a time scaling with $\epsilon \mathcal{N}$ so that microscopic objects take arbitrarily long to be affected by a nearly infinitesimal $\epsilon$, while the collapse process is nearly instantaneous in the limit of large $\mathcal{N}$, even for very small non-unitary perturbations. Moreover, the stable end states of the quantum state reduction are given by the symmetry-breaking pointer states, and Born's rule statistics emerge spontaneously. 


\section{One random variable}
\label{Sec:Ns1RV}
Having a model for DQSR based on SUV for the specific case of a two-state superposition of pointer states, we will now generalize the approach to initial superpositions over $N$ pointer states. Notice the difference between $\mathcal{N}$ (the size of the measurement apparatus) and $N$ (the number of pointer states with nonzero weight in the initial superposition). The generalization can be done in multiple ways, differing in the number of required stochastic variables and the symmetry properties of the non-unitary perturbation.

The mathematically most straightforward extension of the two-state evolution can be found by first rewriting Eq.~\eqref{eq:thetadot} in the form:
\begin{align}
\hbar \, d\theta/dt = \epsilon \mathcal{N} \sin(\theta) \left( \lambda - \cos^2(\theta/2) \right)
\label{Eq:thetadot2}
\end{align}
Here, the random variable $\xi \in U[-1,1]$ was replaced with $\lambda = (\xi+1)/2$, which corresponds to a random variable taken from a uniform distribution on the domain $[0,1]$. This rewriting of the time evolution brings to the fore two important points. First, it makes clear why Born's rule emerges. The relative weights in the two-state superposition are determined at any time by $\theta$, with pointer states corresponding to $\theta=0$ and $\theta=\pi$. If the value of $\lambda$ in Eq.~\eqref{Eq:thetadot2} is lower than $\cos^2(\theta_0/2)$, then the velocity $d\theta/dt$ is negative and the value of $\theta$ will decrease, indicating an evolution towards $\theta=0$. Since $\theta$ decreases, $\lambda - \cos^2(\theta/2)$ will also decrease, and the sign of the velocity never changes (that is, the evolution in Fig.~\ref{fig1} never crosses the separatrix). Thus, for every value of $\lambda$ smaller than  $\cos^2(\theta_0/2)$, the pointer state at $\theta=0$ ends up as the final outcome of the DQSR process.

The probability for finding the state $\ket{1}$ (i.e. $\theta=0$) as the result of the quantum measurement is now understood to equal the probability for the term $\lambda-|\beta_0|^2/(|\alpha_0|^2+|\beta_0|^2)$ to be negative. If $\lambda$ is randomly taken from $U[0,1]$ that probability is $|\beta_0|^2/(|\alpha_0|^2 +|\beta_0|^2)$, in agreement with Born's rule.

Secondly, the set of possible final states and their corresponding probabilities will not change if all diagonal elements of $\hat{G}$ are multiplied by a common factor. Such an overall multiplicative factor would affect the speed with which components evolve during the DQSR process, but not the locations of fixed points or separatrices.

Having identified these characteristics, we can propose a generalization. Consider an initial superposition over $N$ pointer states, written as:
\begin{align}
    \ket{\psi}=\sum_{j=0}^{N-1} {\alpha_j} \ket{j}, ~~~~\text{with} ~~~~ \sum_{j=0}^{N-1}\abs{\alpha_j}^2=1
    \label{eq:N-state}
\end{align}
To avoid imposing normalization at every time step, we again switch to a representation on a higher-dimensional generalization of the Bloch sphere. Introducing angles $\theta_m$ with $m \in \{1, 2, \dots, N-1\}$ describing the relative weights of components, we write:
\begin{align}
    |\alpha_{N-1}| &= \prod_{m=1}^{N-1} \cos\left(\frac{\theta_m}{2}\right)\notag\\    
    |\alpha_{0<j<N-1}| &= \sin\left(\frac{\theta_{j+1}}{2}\right) \prod_{m=1}^{j} \cos\left(\frac{\theta_m}{2}\right)\notag\\
    |\alpha_{0}| &= \sin\left(\frac{\theta_{1}}{2}\right)
    \label{eq:alphadefs}
\end{align}

In direct analogy with the two-state process, we would like the pointer state to correspond to fixed points of the non-linear time evolution in the state-space spanned by the variables $\theta_m$. On the level of the evolution equation, this can is accomplished by having $d\theta_m/dt \propto \sin(\theta_m)$. The flow lines then end at points in phase space where all $\theta_m$ equal either zero or $\pi$, or equivalently at the states $\ket{j}$ (and not superpositions of them). Notice that in fact, the state $\ket{0}$ corresponds to $\theta_1=\pi$, irrespective of the values of $\theta_m$ for $m > 1$, because of the factor $\cos(\theta_1/2)$ appearing in all $|\alpha_j|$ except $|\alpha_{0}|$. Similarly, $\ket{1}$ corresponds to $\theta_1=0$ and $\theta_2=\pi$, regardless of the values of $\theta_m$ for $m > 2$, and so on.

\begin{figure*}
\includegraphics[width=\textwidth]{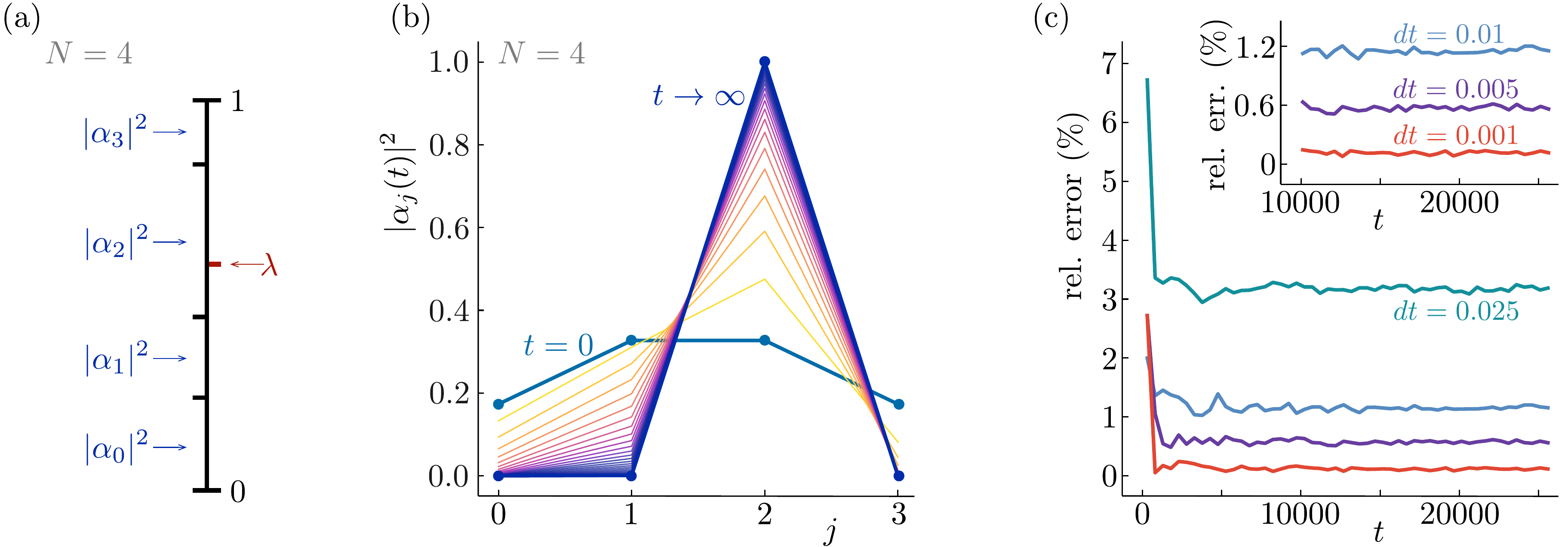}
\caption{Quantum state reduction with one random variable. (a) The line interval $[0,1]$ can be divided into pieces with lengths corresponding to the weights $|\alpha_j|^2$ of pointer states in an initial state wave function. The probability for a stochastic variable $\lambda$ randomly chosen from a uniform distribution on $[0,1]$ to have a value corresponding to the state $\ket{j}$, is then equal to $|\alpha_j|^2$. (b) Example of an initial ($t=0$) state superposed over four pointer states $\ket{j}$, being dynamically reduced according to Eq.~\eqref{Eq:thetadotN}, for a particular randomly selected value of the stochastic variable, to a single measurement outcome at late times ($t\to\infty$). (c) The relative deviation from Born's rule of the obtained distribution of final states, as a function of time for different values of the numerical time step $dt$. The relative error equals the absolute difference between $|\alpha_j|^2$ at the initial time and the fraction of simulations ending in state $\ket{j}$, summed over all $j$. In the continuum limit $dt\to 0$, the agreement with Born's rule can be seen to become exact. These curves are for averages over the stochastic variable starting from the initial state depicted in panel (b). Similar results are obtained both for different initial state configurations, and for initial superpositions over different numbers of pointer states.
}
\label{fig2}
\end{figure*}

Having ensured that the possible endpoints of evolution coincide with the pointer states $\ket{j}$, we need to ensure the emergence of Born's rule. That is, each possible final state $\ket{j}$ should have probability $|\alpha_j|^2$ of being selected by the state dynamics. This can be achieved by noticing that in a normalized state vector, the squared components of the wave function add up to one, so that we can interpret them as the lengths of line segments adding up to a line of total length one, as indicated in Fig.~\ref{fig2}(a). The domain of the random variable $\lambda$ is $[0,1]$, so that the value of $\lambda$ can be indicated along the same line in Fig.~\ref{fig2}(a). The probability for the value of $\lambda$ to lie within the block of size $|\alpha_j|^2$ at $t=0$ is equal to the value of $|\alpha_j|^2$ at $t=0$ itself. If the evolution ends up with the final state $\ket{j}$ whenever $\lambda$ starts out in the the block of size $|\alpha_j|^2$, Born's rule is guaranteed to emerge.

The boundary values of $\lambda$, at which the evolutuion should switch from favouring one final state to another, are defined by: 
\begin{align}
    \lambda &= \sum_{j=0}^{n-1} |\alpha_j|^2 = 1-\prod_{m=1}^{n} \cos^2\left(\frac{\theta_m}{2}\right)
    \label{eq:lambdas}
\end{align}
Notice that these define $N-1$ boundary values, one for each value of $n \in \{1, 2, \dots, N-1\}$. They can equivalently be thought of as defining $N-1$ hypersurfaces or separatrices in the space spanned by the angles $\theta_m$. We will write the $N-1$ relations in Eq.~\eqref{eq:lambdas} as $L_n=0$ with $L_n\equiv1-\prod_{m=1}^{n} \cos^2(\theta_m/2)-\lambda$.

To define the evolution of the state, recall from Eq.~\eqref{eq:alphadefs} that the pointer state $\ket{0}$ corresponds to $\theta_1=\pi$, irrespective of the values of $\theta_m$ for $m > 1$. Repeating the reasoning that led to Born's rule in the two-state dynamics, we would thus like to see that $\theta_1$ increases in time and flows towards $\pi$ whenever $\lambda$ is smaller than the value of $1-\cos^2(\theta_1/2)$ at $t=0$, and opposite otherwise. That is, we should demand $d\theta_1/dt \propto L_1$. 

If $\theta_1$ does evolve to $\pi$, Eq.~\eqref{eq:alphadefs} shows that the remainder of the evolution for the other $\theta_m$ can be ignored, as it does not influence the final state. In the opposite case, of $\theta_1$ evolving to zero, the final state will certainly not be $\ket{0}$. Given that $\theta_1$ will become zero, the final state will be $\ket{1}$ if $\theta_2$ evolves towards $\pi$, and some other state otherwise. In fact, as observed before, the state $\ket{1}$ is realised for $\theta_2=\pi$ regardless of the values of $\theta_m$ for $m > 2$. If we demand $d\theta_2/dt \propto L_2$, we thus end up at the final state $\ket{1}$ if $\lambda$ is smaller than $1-\cos^2(\theta_1/2)\cos^2(\theta_2/2)$, but larger than $1-\cos^2(\theta_1/2)$ at $t=0$, establishing agreement with Born's rule for the second component. Iterating this argument, we find that we should demand $d\theta_n/dt \propto L_n$ for all $n$.

These relations are, however, not sufficient to define the dynamics. We ensured that the hypersurface $L_n=0$ separates regions of opposite sign for the evolution of the parameter $\theta_n$, but we have not yet ascertained that the total evolution comes to a standstill at these hypersurfaces such that the evolution does not cross the newfound separatrix.  In other words, we still need to force $d\theta_n/dt=0$ on all hypersurfaces $L_m$ with $m\neq n$. This can be done without affecting the sign of the evolution anywhere by demanding $d\theta_n/dt \propto \prod_{m\neq n}L_m^2$. Since $L_m$ goes to zero whenever the state state approaches the $m^{\text{th}}$ separatrix, $d\theta_n/dt$ is now guaranteed to go to zero at all separatrices. Moreover, since $L_m^2$ is positive on both sides of the $m^{\text{th}}$ separatrix, the sign of $d\theta_n/dt$ is determined solely by which side of the $n^{\text{th}}$ separatrix the state is on.

Putting everything together, we finally find that the time evolution guaranteeing Born's rule is given by:
\begin{align*}
    \hbar\,\frac{d\theta_n}{dt} &= \epsilon \mathcal{N} \sin(\theta_n) L_n \prod_{m\neq n}L_m^2
\end{align*}
In fact, we can simplify this expression by noticing that just as in the two-state case, a single factor multiplying the time derivative of all angles does not change the fixed points or separatrices, and hence leaves the final states and their probabilities invariant. We thus absorb the common factor $\prod_{m}L_m^2$ in the definition of $\epsilon$, keeping in mind that spontaneous unitarity violations will emerge in the limit $\epsilon\to 0$, and end up with the final expression:
\begin{align}
   \hbar \, \frac{d\theta_n}{dt} &= \epsilon \mathcal{N} \frac{\sin(\theta_n)}{  1-\prod_{m=1}^{n} \cos^2(\theta_m/2)-\lambda }
   \label{Eq:thetadotN}
\end{align}

These equations define a model for DQSR starting from an $N$-state superposition in the initial state. The spontaneous breakdown of unitarity takes place in a time scaling with $\epsilon \mathcal{N}$, so that the collapse process for a vanishingly small non-unitary perturbation is effective only in the thermodynamic limit. Moreover, the stable end states of the quantum state reduction are given by the symmetry-breaking pointer states, and Born's rule statistics emerge spontaneously in the process, using just a single random variable chosen from a state-independent, uniform distribution.

Fig.~\ref{fig2} shows a numerical simulation of the dynamics implied by Eq.~\eqref{Eq:thetadotN}. An example of a single evolution, with one value for the random variable $\lambda$, is displayed in panel \ref{fig2}(b), where DQSR to a single pointer state can be clearly seen. The state is normalized at each time step in order to allow visualization of the time evolution. As argued before, the normalization does not influence the final states obtained in the DQSR process, nor their probability distribution. The statistics of an ensemble of evolutions starting from the same initial state by halting each individual realisation of the dynamics whenever the relative weight of a single component exceeds a threshold value. The corresponding pointer state is then selected as the final state for that particular evolution. The deviations of the statistics from Born's rule are shown in Fig.~\ref{fig2}(c) to converge to zero as their numerical simulation approaches the continuum limit.

\begin{figure*}
\includegraphics[width=\textwidth]{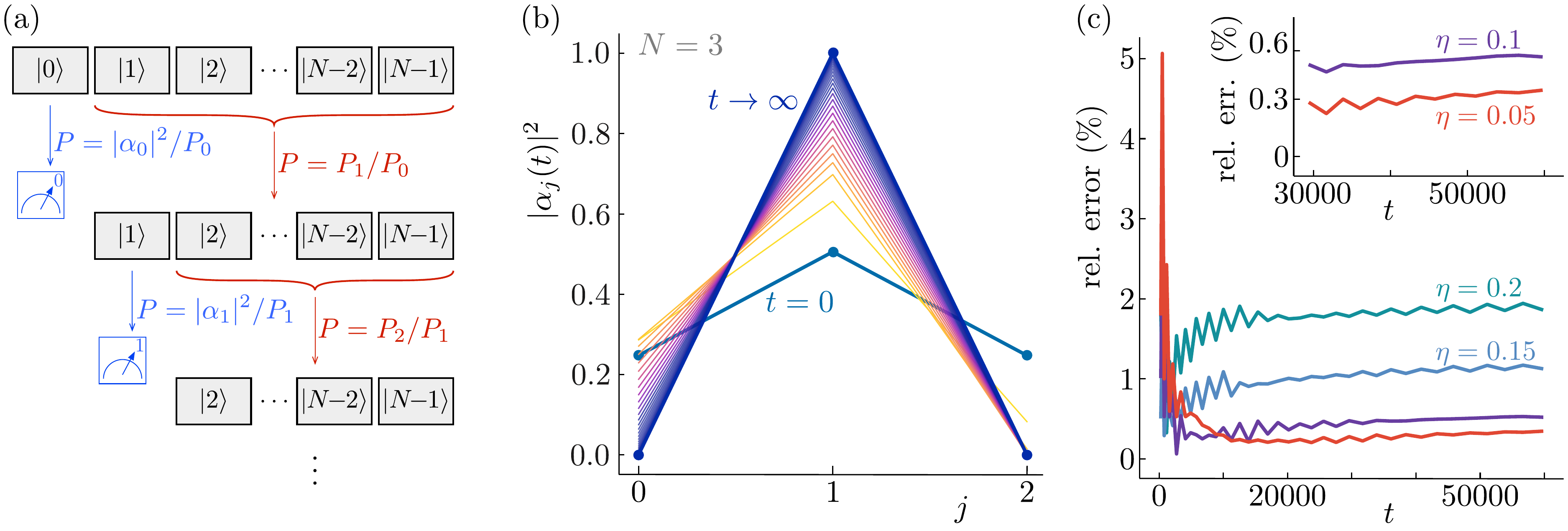}
\caption{Quantum state reduction with $N-1$ random variables. (a) At each stage in the time evolution defined by Eq.~\eqref{Eq:Nrvfinal}, the relative weight of one component of the initial $N$-state superposition evolves to either one or zero. The different stages are separated in time by the proportionality of their evolutions to different powers of the small parameter $\eta$. (b) Example of an initial ($t=0$) state superposed over three pointer states $\ket{j}$, being dynamically reduced according to Eq.~\eqref{Eq:Nrvfinal}, for particular randomly selected values of the stochastic variables, to a single measurement outcome at late times ($t\to\infty$). (c) The relative deviation from Born's rule of the obtained distribution of final states, as a function of time for different values of the small parameter $\eta$. The relative error equals the absolute difference between $|\alpha_j|^2$ at the initial time and the fraction of simulations ending in state $\ket{j}$, summed over all $j$. In the limit of vanishing $\eta$, the agreement with Born's rule can be seen to become exact. These curves are for averages over the stochastic variables starting from the initial state depicted in panel (b). Similar results are obtained both for different initial state configurations, and for initial superpositions over different numbers of pointer states.
}
\label{fig3}
\end{figure*}


\section{Multiple random variables}\label{Sec: N state N-1 RV}
In the previous section, we generalized the description of SUV as a model for DQSR from initial superpositions over two pointer states to an arbitrary number of pointer states in the initial superposition. The generalization based on dividing the $N$-particle phase space into regions of attraction for the $N$ distinct pointer states is mathematically economic because it requires only a single random variable. The final form of the time evolution in Eq.~\eqref{Eq:thetadotN}, however, does not seem to have an obvious interpretation in terms of physical interactions. In this section and the next, we therefore introduce an alternative generalization, which more readily allows for physical interpretation. We first introduce the construction in this section, resulting in a model for DQSR of $N$-state superpositions using $N-1$ random variables. In the next section, we further refine the approach resulting in a model with $\log_2(N)$ random variables, which can be interpreted as components of a continuous field.

Rather than directly dividing the $N$-particle phase space into $N$ domains, we will accomplish the partitioning through a series of binary divisions. The most straightforward way to do this is to first define a time evolution that causes the weight of just one of the pointer states, say $|\alpha_{0}| = \sin(\theta_1/2)$ to become either zero or one:
\begin{align}
    \hbar \, {d\theta_1}/{dt} &= \epsilon \mathcal{N} \sin(\theta_1)\left( \lambda_1-\cos^2(\theta_1/2) \right)
   \label{Eq:thetadotNN1}
\end{align} 
If $\theta_1$ becomes $\pi$, all components $|\alpha_{j}|$ with $j$ larger than one will be zero, and Eq.~\eqref{Eq:thetadotNN1} defines the entire DQSR process. If it evolves to zero, on the other hand, we are left with a superposition over $N-1$ pointer states. We can then define the time evolution for the next component, $|\alpha_{1}| = \sin(\theta_2/2)\cos(\theta_1/2) = \sin(\theta_2/2)$, so that it becomes either zero or one:
\begin{align}
    \hbar \, {d\theta_2}/{dt} &= \eta \epsilon \mathcal{N} \sin(\theta_2)\left( \lambda_2-\cos^2(\theta_2/2) \right)
   \label{Eq:thetadotNN2}
\end{align}
Notice that we introduced a second random variable in this equation. Moreover, to ensure that the dynamics of $|\alpha_{0}|$ is effectively completed before $|\alpha_{1}|$ starts evolving, we introduced the small parameter $\eta$. In the limit $\eta\to 0$, the evolutions of the two components become independent and sequential. 

This procedure can now be iterated, as illustrated in Fig.~\ref{fig3}a, where an $N$-state system undergoes $N-1$ steps with effective two-state evolution. At each level of the partitioning, an independent stochastic component, $\lambda_m$ is introduced, and the evolutions are guaranteed to be independent by scaling their evolution rate with $\eta^m$. We then finally find the complete definition for the dynamics:
\begin{align}
    \hbar \, {d\theta_m}/{dt} &= \eta^m \epsilon \mathcal{N} \sin(\theta_m)\left( \lambda_m-\cos^2(\theta_m/2) \right)
   \label{Eq:thetadotNNall}
\end{align}
Alternatively, the evolution can be specified through the generator $\hat{G}$ acting on the state $\ket{\psi}$ as defined in Eqs.~\eqref{Eq:key} and~\eqref{eq:N-state}. Its diagonal elements $G_j$ are then given by:
\begin{align}
    G_0 &= \eta^0 \left[\frac{|\alpha_{0}|^2-P_1}{P_{0}} - \xi_{0}\right] \notag \\
    {G}_{0<j< N-1} &= \eta^j \left[\frac{|\alpha_{j}|^2-P_{j+1}}{P_{j}} - \xi_{j}\right] \notag \\
    &+ \sum_{m=0}^{j-1} \eta^{m} \left[\xi_{m}-\frac{|\alpha_{m}|^2-P_{m+1}}{P_{m}}\right] \notag \\
    {G}_{N-1} &= \sum_{m=0}^{N-2} \eta^{m} \left[\xi_{m}-\frac{|\alpha_{m}|^2-P_{m+1}}{P_{m}}\right]
    \label{Eq:Nrvfinal}
\end{align}
Here, we defined $P_m = \sum_{j=m}^{N-1} |\alpha_j|^2$, and we reintroduced the random variables $\xi_m=2\lambda_m-1$ sampled from $U[-1,1]$. Just as in Eqs.~\eqref{Eq:key} and~\eqref{2 state eq1}, the time evolution defined by Eq.~\eqref{Eq:Nrvfinal} is not norm-conserving. As before, this is not a problem since it does not affect any physical expectation values~\cite{Mertens_PRA_21}. In numerical simulations of the dynamics, however, it may be convenient to normalise the state either at the end of the calculation, or after every time step. The resulting final state is not affected by this choice.

Notice there is an (arbitrary) hierarchical structure built into the time evolution of Eq.~\eqref{Eq:Nrvfinal}. The time evolution first determines whether pointer state $\ket{0}$ will end up as the final state of the measurement process. This happens with the probability as found in the two-state evolution of Sec.~\ref{sec:1},  $\sin^2(\theta_1/2) = |\alpha_{0}|^2$, in agreement with Born's rule. If $\ket{0}$ is not the final state, the evolution continues, and determines whether pointer state $\ket{1}$ will be the final state. This happens with probability $\sin^2(\theta_2/2)$, but because it can only happen if $\ket{0}$ did not dominate, the total probability for state $\ket{1}$ to be the final state is $\cos^2(\theta_1/2)\sin^2(\theta_2/2)$, again in agreement with Born's rule. 

Continuing this way, the probabilities for all pointer states are seen to agree with Born's rule. This process only works however, if the hierarchy is strictly obeyed and the evolution of $\ket{0}$ is finalised before that of $\ket{1}$ begins, and so on. This is true in the limit $\eta\to 0$, but for finite $\eta$ the final state probabilities will deviate $O(\eta)$ from Born's rule.

The hierarchy introduced by the powers of $\eta$ that is necessary to establish Born's rule implies an arbitrary choice for which pointer state is associated with which power of $\eta$. Although this choice does not influence the final state statistics, it does determine the finite-time dynamics and there is no clear physical reason to favour one choice over any other. In the next section, we will introduce an alternative hierarchy that results in a symmetric form of the time evolution generator, as well as a greatly reduced number of stochastic variables. 

Despite these caveats, Eq.~\eqref{Eq:thetadotNNall}, or equivalently, Eq.~\eqref{Eq:Nrvfinal}, does define a model for DQSR starting from an $N$-state superposition in the initial state. The spontaneous breakdown of unitarity now takes place in a time scaling with $\eta^{N-2} \epsilon \mathcal{N}$. As in the previous section, the collapse process is effective for a vanishingly small non-unitary perturbation in the thermodynamic limit $\mathcal{N}\to\infty$ and the stable end states are given by symmetry-breaking pointer states. This time, Born's rule statistics emerge spontaneously using $N$ independent random variables, each of which is chosen from a state-independent, uniform distribution.

The emergence of stable pointer states and Born's rule can be verified numerically, as shown in figure~\ref{fig3}. Panel \ref{fig3}(b) illustrates an individual instance of the time evolution generated by Eq.~\eqref{Eq:thetadotNNall}. The deviations of the statistics from Born's rule obtained from the ensemble average over many iterations are shown in figure~\ref{fig3}(c) to converge to zero as the hierarchy parameter $\eta$ decreases after approaching the continuum limit. Further details of the numerical simulations may be found in Appendix.\ref{app:numerics}.

\begin{figure*}
\includegraphics[width=\textwidth]{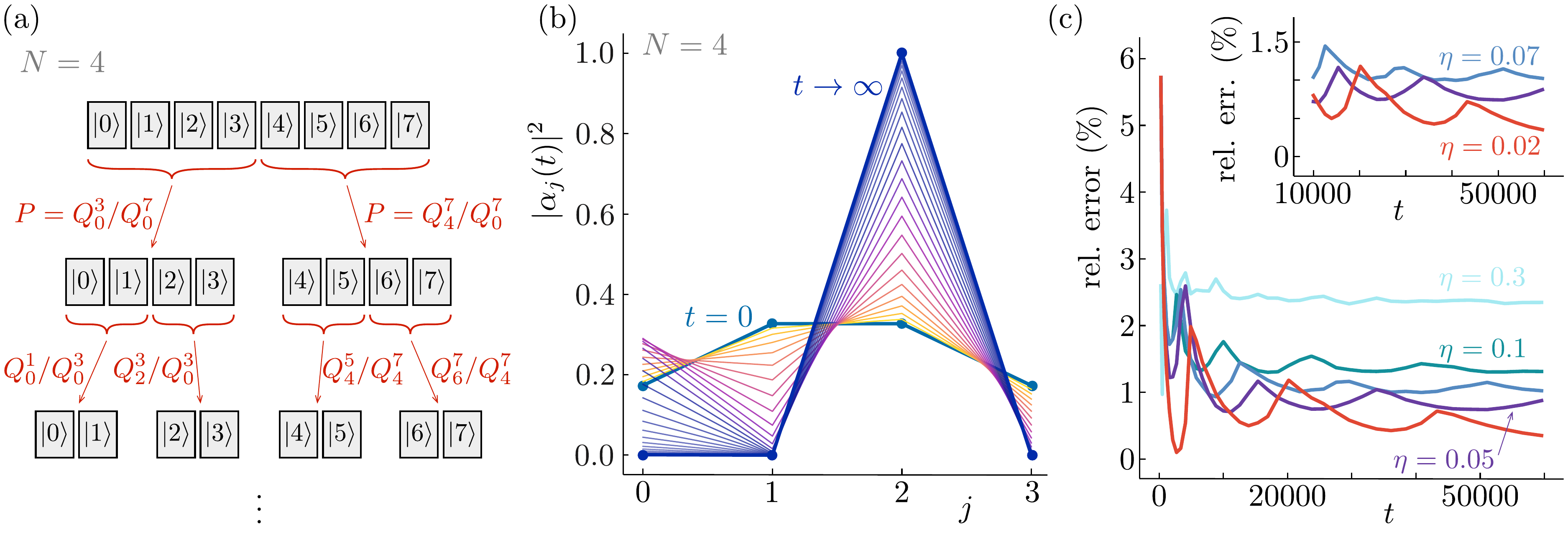}
\caption{Quantum state reduction with $\log_2(N)-1$ random variables. (a) At each stage in the time evolution defined by Eq.~\eqref{N state eq2}, the combined relative weight of one half of the components of the initial $N$-state superposition evolves to either one or zero. At each stage a more fine-grained division of the initial pointer states is used. The different stages are separated in time by the proportionality of their evolutions to different powers of the small parameter $\eta$. (b) Example of an initial ($t=0$) state superposed over four pointer states $\ket{j}$, being dynamically reduced according to Eq.~\eqref{N state eq2}, for particular randomly selected values of the stochastic variables, to a single measurement outcome at late times ($t\to\infty$). (c) The relative deviation from Born's rule of the obtained distribution of final states, as a function of time for different values of the small parameter $\eta$. The relative error equals the absolute difference between $|\alpha_j|^2$ at the initial time and the fraction of simulations ending in state $\ket{j}$, summed over all $j$. In the limit of vanishing $\eta$, the agreement with Born's rule can be seen to become exact. These curves are for averages over the stochastic variables starting from the initial state depicted in panel (b). Similar results are obtained both for different initial state configurations, and for initial superpositions over different numbers of pointer states.
}
\label{fig4}
\end{figure*}

  
\section{A natural hierarchy\label{Sec:NslogNRV}}
We will now show that the series of sequential bipartite collapse evolutions used in the previous section to construct a DQSR model based on spontaneous unitarity violations, can be organised in an alternative way. This will both be more mathematically efficient, using only $\log_2{N}$ random variables rather than $N-1$, and more physically appealing, as it yields a more symmetric form of the generator for time evolution that allows a natural continuum limit.

We will again consider the initial state of Eq.~\eqref{eq:N-state} and construct a sequence of binary collapse processes. Rather than having each process determine the fate of a single pointer state, however, each stage of the evolution suppresses the weight of half of all pointer states to zero. As shown in figure~\ref{fig4}(a), the first stage suppresses either the weight of states $\ket{j}$ with $j=0 \dots N/2-1$, or that of the states with $j=N/2 \dots N-1$. In the second stage, each of these blocks has half of their states suppressed to zero weight, and subsequent stages likewise divide each of the blocks created by their predecessor.

As before, each stage in this sequential process utilizes a separate, independent random variable $\xi_p \in [-1,1]$, and has its time evolution scaled by a different power of the small parameter $\eta$. Because all pointer states are involved at all stages, a total of $\log_2(N)$ partitions suffice to single out a final state for the measurement process starting from a superposition of $N$ pointer states. 

The form of the time evolution for this sequence of bipartite evolutions is most easily formulated directly in terms of the generator $\hat{G}$, rather than on the generalised Bloch sphere. To ensure the emergence of Born's rule, the combined squared weights of half of all pointer states evolves to either zero or one during each of the stages sketched in Fig.~\ref{fig4}(a), but the relative weights within each evolving half are not affected. We can thus directly generalise the result of Eq.~\eqref{2 state eq1} to write for the first stage:
\begin{align}
\hat{G}^{(0)} &= \sum_{j=0}^{N/2-1} \ket{j}\left[\frac{Q_{0}^{N/2-1} - Q_{N/2}^{N-1}}{Q_0^{N-1}} -\xi_0\right]\bra{j} \notag \\
&~~~~~ + \sum_{j=N/2}^{N-1} \ket{j}\left[\xi_0 - \frac{Q_{0}^{N/2-1} - Q_{N/2}^{N-1}}{Q_0^{N-1}}\right]\bra{j}
\label{N state eq1}
\end{align}
Here, we defined $Q_m^n=\sum_{j=m}^{n} |\alpha_j|^2$, and the total generator is divided into stages as $\hat{G} = \sum_{p=0}^{\log_2(N)-1} \hat{G}^{(p)}$, with the power of $\eta$ increasing in each consecutive stage (here, $\hat{G}^{(0)}$ implicitly includes a factor $\eta^0$).

Generalizing directly to the full expression, we find:
\begin{align}
\hat{G}^{(p)} &= \sum_{j=0}^{N-1} \ket{j} \eta^{p} \Theta(j,p) \left[\sum_{j'=0}^{N-1} \Theta(j',p) \frac{|\alpha_{j'}|^2}{Q_0^{N-1}} -\xi_p\right]\bra{j} \notag \\
&\text{with} ~~~ \Theta(j,p) = (-1)^{\big{\lfloor{j 2^{p+1} / N}\rfloor}}
\label{N state eq2}
\end{align}
Here $\lfloor{z}\rfloor$ is the floor of $z$, which equals the largest integer smaller than or equal to $z$. The value of $\Theta(j,p)$ is then either $+1$ or $-1$, and this function partitions the pointer states at each stage of the evolution.

The independence of subsequent stages in the collapse process is guaranteed by $\eta$ being a small parameter, as in the previous section. Since Born's rule was shown to emerge in the two-state process of Eq.~\eqref{2 state eq1}, it is also guaranteed to emerge from Eq.~\eqref{N state eq2} in the limit of vanishing $\eta$. For finite values of $\eta$, deviations from Born's rule of order $\eta$ will occur.

Equation~\eqref{N state eq2} is one of the main results of this article. It defines a model for DQSR starting from an $N$-state superposition in the initial state. The spontaneous breakdown of unitarity takes place in a time scaling with $\eta^{\log_2(N)} \epsilon \mathcal{N}$, so that the collapse process is effective for a vanishingly small non-unitary perturbation in the thermodynamic limit $\mathcal{N}\to\infty$. The stable end states of the quantum state reduction are given by the symmetry-broken pointer states, and Born's rule statistics emerge spontaneously in the process, using $\log_2(N)$ independent random variables, each of which is chosen from a state-independent, uniform distribution. Moreover, despite the hierarchy of the collapse process, the form of Eq.~\eqref{N state eq2} is symmetric in the sense that all pointer states evolve during all stages of the DQSR process. 

The division of pointer states into two groups at each stage can be interpreted as a stepwise fine-graining of the measurement outcome. Since pointer states correspond to classical symmetry-broken states of matter, they differ in the value or direction of an order parameter~\cite{Wezel_SSBlecturenotes,Wezel10}. For an actual pointer along a dial, for example, this could be the position of the tip of the pointer. This means there is a natural ordering of pointer states, in the order parameter space. The states of an actual pointer, for example, could be ordered in real space, going from one end of the dial to the other. Within this natural ordering, the first stage of the DQSR process described by Eq.~\eqref{N state eq2} then suppresses one connected set of pointer states, establishing that the measurement outcome will fall within the remaining half. The second stage suppresses a connected section of the remaining states and establishes the quarter of all initial states among which the final state will fall. Continuing this way, each consecutive stage of the process gives a more fine-grained set of candidates for the final state. This interpretation of fine-graining in an order parameter space suggests a natural continuum limit for Eq.~\eqref{N state eq2}, which we will explore in the following section.

As in previous sections, the emergence of stable pointer states and Born's rule can again be verified numerically, as shown in Fig.~\ref{fig4}. Panel \ref{fig4}(b) illustrates an individual instance of the time evolution generated by Eq.~\eqref{N state eq2}. The deviations of the statistics from Born's rule obtained from the ensemble average over many iterations are shown in fig.~\ref{fig4}(c) to converge to zero as the hierarchy parameter $\eta$ decreases after approaching the continuum limit. Further details of the numerical simulations may be found in Appendix.\ref{app:numerics}.


\section{Towards a random field\label{Sec:field}}
The final form of the DQSR process with $\log_2(N)$ random variables in Eq.~\eqref{N state eq2} suggests a natural generalization to a model for quantum measurement with the initial state superposed over a continuous set of states. Without loss of generality, consider a line segment parameterized by the coordinate $x \in [0,1]$. The initial state is now:
\begin{align}
\ket{\psi}=\int_0^1  dx \,\psi(x) \ket{x} ~~~ \text{with} ~~~ \int_0^1  dx  \,\abs{\psi(x)}^2=1
\label{eq:ContState}
\end{align}
Taking the discrete pointer states $\ket{j}$ of the previous section to lie within the continuous interval parameterized by $x$ and taking the continuum limit $N\to\infty$ after identifying $x=j/N$, the contribution to the time evolution generator at stage $p$ becomes:
\begin{align}
\hat{G}^{(p)} &= \int_{0}^{1}dx\, \ket{x} \eta^{p} \theta(x,p) ~\times \notag \\
&~~~~~~~~~\left[\int_{0}^{1}dx'\, \theta(x',p) \frac{|\psi(x')|^2}{Q} -\xi(p) \right]\bra{x}
\label{contG}
\end{align}
Here, we introduced the generally time-dependent norm $Q(t)=\int_0^1 dx\, |\psi(x,t)|^2$ as well as the continuum version of the sign distribution function on the interval $[0,1]$, given by $\theta(x,p) = (-1)^{\big{\lfloor{x 2^{p+1}}\rfloor}}$. The full generator is given by $\hat{G}=\sum_{p=0}^{\gamma} \hat{G}^{(p)}$, with $\gamma$ an ultraviolet cutoff. 

The full time evolution generator can be cast in a more suggestive form by defining:
\begin{align}
    \hat{G} \ket{\psi} = \int_0^1 dx\, G(x) \psi(x) \ket{x}
\end{align}
The non-linear components of $\hat{G}$ are then given by:
\begin{align}
G(x)&= \Lambda(x) + \int_0^1 dx'\, \frac{\abs{\psi(x')}^2}{Q} \Pi(x,x') \notag \\
&= \Lambda(x) + \left<\hat{\Pi}(x)\right>.
\label{Eq:G_fin}
\end{align}
The expectation value $\langle\hat{\Pi}(x)\rangle$ resembles a spatial propagator with elements $\Pi(x,x')=\sum_{0}^{\gamma} \eta^{p} \, \theta(x,p)\theta(x',p)$, while $\Lambda(x)=-\sum_{0}^{\gamma} \eta^{p}\xi_p\theta(x,p)$ represents the value at location $x$ of a random field on the line segment $[0,1]$. Because the stages labeled by $p$ represent different levels of fine-graining in the $x$-space resolution of the final pointer state, the ultra-violet cut-off $\gamma$ also defines a minimum separation for which points along the $[0,1]$ line segment can be resolved. If the pointer states break a symmetry corresponding to an order parameter labeled by a real-space coordinate (such as an actual pointer along a dial), the ultraviolet cutoff could for example be set by the Planck length. Measurement outcomes can then only ever be resolved down to Planck length precision, and the random field $\Lambda(x)$ takes independent random values on positions separated by a Planck length.


\section{Discussions and Conclusions\label{Sec:conclusions}}
In conclusion, we constructed several models for dynamic quantum state reduction based on the idea that the time inversion symmetry underlying unitarity in quantum dynamics can be spontaneously broken, like any other symmetry in nature. Although it has been known for some time that the unitary dynamics of Schr\"odinger's equation is unstable in the thermodynamic limit~\cite{Wezel_2008, Wezel10}, a concrete model for the unitarity-breaking time evolution starting from a generic initial state and obeying all requirements for a model of quantum measurement was still lacking. Here, we showed that the measurement dynamics previously proposed for an initial superposition over two pointer states~\cite{Mertens_PRA_21} can be generalized to arbitrary initial states in several ways, which differ in the way Born's rule emerges during the measurement process. Note, however, in all the generalizations considered, Born's rule emerges by construction and not as a result of imposing it.

We first considered a mathematically straightforward generalization, in which just a single random variable chosen from a flat, uniform distribution leads to precisely Born's rule for an initial superposition of an arbitrary finite number of pointer states. This model, however, does not have a straightforward physical interpretation.

Next, we constructed a generalization using as many random variables as there are pointer states (minus one) in the initial superposition. The emergence of Born's rule in this model relies on the presence of separate stages in the measurement dynamics and is perfect only in the limit of vanishing overlap between these stages. Moreover, the model requires the introduction of an arbitrary hierarchy among the pointer states.

The final generalization we introduced removes the arbitrary hierarchy and replaces it with a natural ordering of the pointer states interpreted as symmetry-breaking states with a macroscopic order parameter. This way, only $\log_2(N)$ random variables are required to model the dynamical quantum state reduction of an initial superposition over $N$ pointer states. Moreover, the final generator for time evolution in the model has a natural continuum limit, which can be interpreted in terms of a random field in real space and an expectation value resembling a real-space propagator.

The final model for the state reduction dynamics meets all requirements for a model of quantum measurement: its origin in a theory for spontaneous unitarity violation implies that it has negligible effect on the microscopic scale of elementary particles, even though it dominates the behavior of macroscopic, everyday objects and causes them to collapse almost instantaneously. The final states in that collapse process are the symmetry-breaking pointer states that we associate with real-world measurement machines, and after one of them has been selected in the stochastic measurement dynamics, it remains stable. Finally, the probability of finding any particular final state is given by Born's rule, which emerges spontaneously without being used, assumed, or imposed in the definition of the stochastic field. The obtained dynamics does not contradict the experimental observation of Bell inequality violations, as the stochastic noise term acts non-locally on the quantum dynamics. That is, the described dynamics is of the total, extended and entangled quantum state as a whole, and does not employ any of the local hidden variables that are ruled out by Bell tests.

The models presented here explicitly demonstrate the possibility of spontaneous unitarity violations giving rise to DQSR dynamics in a way that obeys all basic requirements for a theory of quantum measurement. The models introduced are non-relativistic and can be extended in several directions, including for example by formulating a field theory in Fock space, or by generalizing the basis of sign functions appearing in the continuum model. Furthermore, it remains to be established whether or not the types of models for spontaneous unitarity violation introduced here allow for superluminal communication. Previous criteria for avoiding non-causal dynamics, by requiring a quantum dynamical semigroup with linear dynamics \cite{Gisin:1989sx,Bassi2015}, were derived for ensemble averages of white-noise driven Markovian models and do not necessarily apply here. Notice that for specific situations in which the noise dynamics is appreciably faster than the quantum state dynamics, an effective Markovian limit with linear master equations, may be achieved by temporal coarse-graining, also called multi-scale noise homogenization, which rules out superluminal signalling in those regimes~\cite{mukherjee2023colored}. We leave the study of these questions in more general situations for future research, and hope the present work will inspire and lay the foundation for further proposals of dynamic quantum state reduction based on spontaneous unitarity violation. These may find application in describing the dynamics of (quantum) phase transitions~\cite{WezelBerry, Wezel_SSBlecturenotes} as well as quantum measurement, yield testable experimental predictions~\cite{wezel_oosterkamp_2011}, and generally shed new light on the crossover regime separating Schr\"odinger from Newtonian dynamics.


\subsection*{Acknowledgement}
The authors gratefully acknowledge stimulating discussions with J. Zaanen on the role of two-fold partitions in the time evolution operator. A.M. also acknowledges supporting discussions with Parameshwari Devi.


\appendix
\counterwithin{figure}{section}

\section{Numerical simulations\label{app:numerics}}
In this appendix we describe the numerical simulations leading to Figs.~\ref{fig2}(b, c), \ref{fig3}(b, c) and~\ref{fig4}(b, c). Convergence in the numerical integration of Eqs.~\eqref{Eq:thetadotN}, \eqref{Eq:Nrvfinal}, and \eqref{N state eq2} is obtained using sufficiently small time steps $dt$. In Fig.~\ref{fig2}(c) we show convergence to Born rule statistics for decreasing value of the time step. The dynamics defined in Secs.~\ref{Sec: N state N-1 RV} and~\ref{Sec:NslogNRV} additionally require a small hierarchical parameter $\eta$. For any given value of $\eta$ the size of $dt$ was adjusted to ensure convergent results, with lower values of $\eta$ requiring smaller time steps. Therefore, in Fig.~\ref{fig3}(c), the values $\eta=0.05$ and $dt=0.005$ were used, while for other values of $\eta$ taking $dt=0.01$ sufficed. The results in Fig.~\ref{fig4}(c) used $dt=0.01$ for all cases except for $\eta=0.05$ and $\eta=0.02$, which both utilized $dt=0.005$.

To recover Born rule statistics, a numerical average must be taken over a dense and uniform set of values for the stochastic variable. The results in Fig.~\ref{fig2}(c) represent averages over $100$ to approximately $25000$ values for the stochastic variable, while up to $60000$ values were sampled in the creation of Figs.~\ref{fig3}(c) and~\ref{fig4}(c).

\section{Continuum distributions\label{app:numerics2}}
\begin{figure}
\includegraphics[width=\columnwidth]{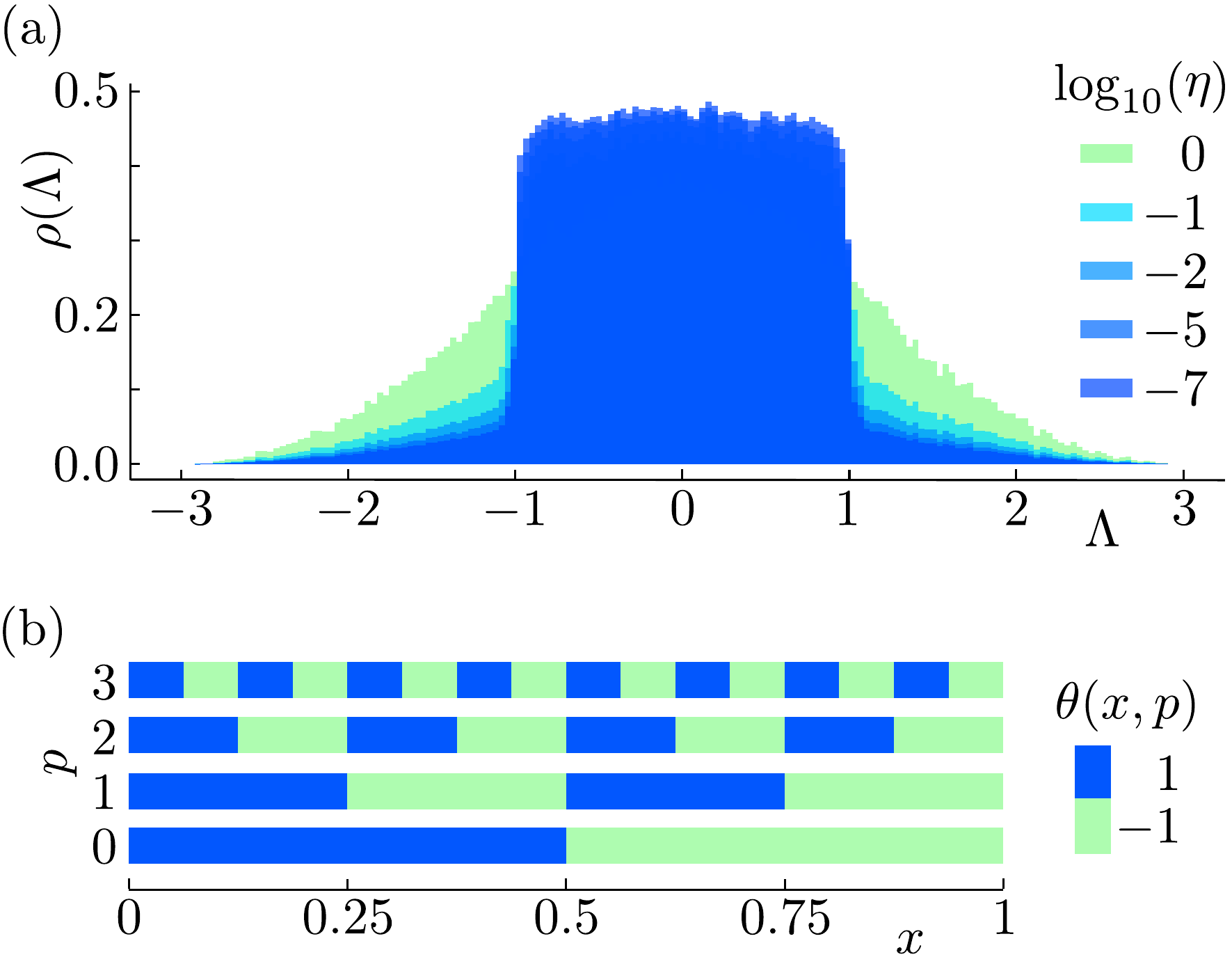}
\caption{(a) The probability distribution function for the random value $\Lambda(x)$, for arbitrary $x$, obtained by numerically averaging over $50000$ randomly selected values for the stochastic variables $\xi_p$. The results for different values of the small parameter $\eta$ converge to a uniform distribution on the interval $[-1,1]$ for vanishing $\eta$. (b) Schematic depiction of the function $\theta(x,p)$, for the continuous variable $x\in[0,1]$ and $p$ discrete.}
\label{figApp}
\end{figure}

In this appendix, we discuss the functions $\Theta(x,p)$ and $\Lambda(x)$ emerging in the continuum theory of Sec.~\ref{Sec:field}. The sign distribution function $\Theta(x,p)$ is defined as $(-1)^{\lfloor{x 2^{p+1} }\rfloor}$ with $x\in[0,1]$. It shown for the first four values of the discrete parameter $p$ in Fig.~\ref{figApp}(b). For any given value of $p$, the function $\Theta(x,p)$ is a square wave, with values alternating between $1$ and $-1$. Notice that any real function on a discrete lattice can be decomposed into these square wave components, much like a Fourier decomposition. To decompose continuous functions, a regularization of the limiting function at $p\to\infty$ will be required.

Finally, the stochastic field $\Lambda(x)$ was defined in Sec.~\ref{Sec:field} as $-\sum_{0}^{\gamma} \eta^{p}\xi_p\theta(x,p)$. The probability density function for $\Lambda(x)$ will be independent of $x$, since it is given by a sum over stochastic variables with coefficients that differing by at most a sign. The probability density function resulting from a numerical evaluation of the sum for $50000$ samples of the random parameters is displayed in Fig.~\ref{figApp}(a), for an arbitrary value of $x$. It corresponds to a type of truncated Gaussian-like distribution for large values of $\eta$, while converging to uniform distribution with tapering edges for smaller values of $\eta$. The tapering at the edges is suppressed as $\eta$ is increased, and the probability density function approaching a true uniform distribution, $U[-1,1]$, as $\eta$ approaches zero. 

\section{Spontaneous Symmetry Breaking \label{app:SSB}}
In this appendix, we review some of the central concepts in the theory of spontaneous symmetry breaking and summarise their use in the models of spontaneous unitarity violation introduced in the main text. For a more detailed and extensive discussion of the physics of spontaneous symmetry breaking, see Ref.~\cite{Wezel_SSBlecturenotes}.

Spontaneous symmetry breaking refers to the situation in which the Hamiltonian governing a system possesses a symmetry, but the actually realised state of the system has a lower symmetry. The `spontaneous' refers to the fact that the symmetry breaking is unavoidable in practice, and that the way in which the symmetry is broken is unpredictable for all practical purposes. For concreteness, we briefly discuss the example of a harmonic crystal, but all concepts apply equally to any system spontaneously breaking a symmetry.

\subsection{The harmonic crystal}
The Hamiltonian describing a harmonic crystal is:
\begin{align}
\hat{H} = \sum_j \frac{\hat{P}^2_j}{2m} + \sum_{\langle i,j\rangle} \frac{1}{2}m \omega^2 \left(\hat{X}_i-\hat{X}_{j}\right)^2.
\label{eq:xtal}
\end{align}
Here, $i$ and $j$ label neighbouring sites of an atomic lattice in which $m$ is the atomic mass and $\omega$ the natural frequency of the (effective) harmonic forces between neighbouring atoms. Both the assumption of a short-ranged interaction potential and that of its harmonic nature can be straightforwardly relaxed in the following. 

The harmonic crystal is symmetric under (global) translations of all of its atoms. Such translations are generated by the total momentum operator $\hat{P}_{\text{tot}}=\sum_j \hat{P}_j$. Because this operator commutes with the Hamiltonian, all eigenstates of $\hat{H}$ are simultaneously eigenstates of $\hat{P}_{\text{tot}}$, which are plane wave states with fully delocalised centre of mass. That is, all eigenstates of $\hat{H}$ respect its translational symmetry, and not the localised states we would expect to find for a macroscopic crystal.

The Fourier transform of Eq.~\eqref{eq:xtal} can be written as:
\begin{align}
\hat{H} = \frac{\hat{P}_{\text{tot}}}{2mN} + \sum_{k\neq 0} \hat{H}_k.
\label{eq:xtalk}
\end{align}
Here, $N$ is the number of atoms in the harmonic crystal, and $k$ denotes the internal crystal momentum. Since we are interested in the global properties of the crystal, we will ignore $\hat{H}_k$ from here on, except for noting that its eigenvalues are all strictly positive and greater than $E_{\text{int}} = \hbar \omega / N^{1/d}$, with $d$ the number of spatial dimensions. At energies or temperatures lower than $E_{\text{int}}$, therefore, the first collective term of Eq.~\eqref{eq:xtalk} dominates. 

The form of the Hamiltonian in Eq.~\eqref{eq:xtalk} clearly shows that the ground state is non-degenerate and has total momentum $\hat{P}_{\text{tot}}=0$. Excitations with non-zero total momentum (up to $P_{\text{tot}}\sim\sqrt{N}$) are separated from the ground state by energies of order $1/N$. This so-called tower of low energy states becomes degenerate with the ground state in the \emph{thermodynamic limit} $N\to \infty$. In that limit, superpositions of total momentum states are also ground states of $\hat{H}$, and it becomes possible for a wave packet to be formed in which the crystal has a localised centre of mass and breaks translational symmetry. For the more physically relevant case in which $N$ is large but not infinite, forcing the crystal into a symmetry-breaking, localised state requires the application of an external force:
\begin{align}
\hat{H}_{\text{SB}} = \frac{\hat{P}_{\text{tot}}}{2mN} + \epsilon N \left(\hat{X}_{\text{com}} - x_0\right)^2.
\label{eq:xtalSB}
\end{align}
Here, $\hat{X}_{\text{com}}$ is the operator for the centre of mass position, $x_0$ is the centre of the externally applied potential, and $\epsilon$ is its strength. The factor $N$ multiplying $\epsilon$ is required for the energy to be extensive, and signals the fact that the applied potential couples to an \emph{order parameter} of the harmonic crystal~\cite{vanwezelAmJPhys}. It is straightforwardly shown that the non-degenerate ground state $\ket{\psi_{\text{gs}}}$ of this Hamiltonian is a Gaussian wave function with the limiting behaviour:
\begin{align}
    \lim_{{N}\to \infty} \lim_{\epsilon\to 0} \ket{\psi_{\text{gs}}} &= \ket{P_{\text{tot}}=0} \notag \\
    \lim_{\epsilon\to 0} \lim_{{N}\to \infty}  \ket{\psi_{\text{gs}}} &= \ket{X_{\text{com}}=x_0}. 
    \label{eq:limits}
\end{align}
That is, if there is no externally applied potential whatsoever, the ground state of the crystal is fully delocalised and symmetric. If there is even an infinitesimally small (but non-zero) perturbation $\epsilon$, however, the crystal ground state is a fully localised symmetry-broken state in the thermodynamic limit. 

Of course, neither of the limits in Eq.~\eqref{eq:limits} are ever realised in nature. What the non-commuting (or \emph{singular}) limits signal is a \emph{diverging susceptibility} of the crystal to symmetry-breaking perturbations. That is, for large crystals consisting of say $N=10^{23}$ atoms, the potential required to force it into a symmetry-broken configuration is of the order of $1/N$, which makes it so small as to be completely beyond the reach of anything we can ever hope to detect, let alone control. For all practical purposes therefore, there will always be some potential or perturbation in any experiment or physical situation that renders the ground states of human-sized harmonic crystals fully localised. Because the localisation is unavoidable, and because the localisation centre $x_0$ is in practice immeasurable, unpredictable, and uncontrollable, we say that symmetry-breaking localisation of the crystal is \emph{spontaneous}.

Notice that the symmetry breaking behaviour \emph{emerges} as the thermodynamic limit is approached. Microscopic harmonic crystals consisting of only a few atoms will not be spontaneously localised, and in fact the extremely weak perturbations that suffice to localise macroscopic crystals will not have more than an undetectably small and negligible effect on microscopic systems. 

Furthermore, the emergent localisation is \emph{universal}, in the sense that the precise shape and strength of the localising potential are irrelevant to the final localised state. Only symmetry-breaking perturbations coupling to the order parameter (i.e. localising the crystal) will have any effect at vanishing strength, and all symmetry breaking perturbations lead to the same type of completely localised ground state for the macroscopic crystal.

\subsection{Spontaneous unitarity breaking}
As shown in Refs.~\cite{vanwezelprb,Wezel10}, the symmetries underlying the unitarity of quantum mechanical time evolution can be spontaneously broken in the same way that any other symmetries of nature are spontaneously broken. That is, the same tower of states with energies vanishing in the thermodynamic limit that allows systems to spontaneously break any regular symmetry \emph{additionally} allows such systems to avoid the unitarity time evolution dictated by Schr\"odinger's equation.

As in the case of regular symmetry breaking, any realistic system of large but finite size will require a non-zero perturbation to affect the breaking of unitarity. In this case, the perturbation must cause non-unitary evolution, and thus corresponds to a non-Hermitian addition to the Hamiltonian: $\hat{H}_{\text{SUV}}=\hat{H} + i \epsilon \hat{G}$. Here, $\hat{H}$ is the Hamiltonian for a system with a spontaneously broken regular symmetry, $\epsilon$ is the strength of the non-unitary perturbation, and $\hat{G}$ is a Hermitian operator coupling to the order parameter of $\hat{H}$. For example, in the case of a harmonic crystal we may consider:
\begin{align}
\hat{H}_{\text{SUV}}=\frac{\hat{P}^2_{\text{tot}}}{2mN} + i \epsilon N \left(\hat{X}_{\text{com}} - x_0\right)^2.
\label{eq:XtalSUVdef}
\end{align}
Here $\epsilon$ is the strength of the non-unitary perturbation, and the factor $N$ again arises from the coupling to the order parameter~\cite{vanwezelAmJPhys}.

The unitarity breaking field causes the emergence of a singular limit analogous to that of Eq.~\eqref{eq:limits}, but now in the time evolution of a given initial state rather than in the definition of the equilibrium ground state:
\begin{align}
\label{eq:XtalSUV}
    \lim_{{N}\to \infty} \lim_{\epsilon\to 0} e^{-\frac{i}{\hbar}t \hat{H}_{\text{SUV}}} \ket{P_{\text{tot}}=0} &= \ket{P_{\text{tot}}=0} \\
    \lim_{\epsilon\to 0} \lim_{{N}\to \infty}  e^{-\frac{i}{\hbar}t \hat{H}_{\text{SUV}}} \ket{P_{\text{tot}}=0} &= \ket{X_{\text{com}}=x_0} ~~~ \forall t>0. \notag
\end{align}
That is, in the complete absence of any non-unitary perturbations whatsoever, the symmetric initial state is stable under the time evolution generated by Schr\"odinger's equation. Importantly, this remains true even in the presence of a potential of the form of Eq.~\eqref{eq:xtalSB}, which breaks the spatial translation symmetry but not the unitarity of time evolution~\cite{WezelBerry}. 

As before, neither of the limits in Eq.~\eqref{eq:XtalSUV} needs to actually be realised in any realistic setting. What the formal existence of these non-commuting, \emph{singular} limits signal, is a \emph{diverging susceptibility} of the crystal to unitarity-breaking perturbations. That is, for large crystals consisting of say $N=10^{23}$ atoms, the perturbation required for it to evolve into a symmetry-broken configuration is of the order of $1/N$, which makes it so small as to be completely beyond the reach of anything we can ever hope to detect, let alone control. For all practical purposes therefore, there will always be some potential or perturbation in any experiment or physical situation that makes it impossible for human-sized harmonic crystals to avoid being localised as a function of time, even if it starts out from a delocalised initial state. Because the evolution towards localisation is unavoidable, and because the localisation centre $x_0$ is in practice unpredictable and uncontrollable, the unitarity of the time evolution may be said to be violated \emph{spontaneously}.

Notice that the breakdown of unitarity \emph{emerges} as the thermodynamic limit is approached. Microscopic harmonic crystals consisting of only a few atoms will not spontaneously evolve away from a delocalised state, and in fact the extremely weak perturbations that suffice to localise macroscopic crystals will take longer than the age of the universe to have a detectable effect on the evolution of microscopic systems. 

Furthermore, the emergent localisation is \emph{universal}, in the sense that the precise shape and strength of the localising potential are irrelevant to the final localised state. Only unitarity-breaking perturbations coupling to the order parameter (i.e. localising the crystal) will have any effect at vanishing strength, and all unitarity breaking perturbations cause evolution towards the same type of localised state.

\subsection{Quantum measurement}
For spontaneous unitarity violations to explain quantum measurement, the non-unitary perturbation in Eq.~\eqref{eq:XtalSUVdef} is not sufficient. As shown in Refs.~\cite{Wezel10, Mertens_PRA_21, Mertens22,Lotte}, Born's rule can emerge from non-unitary dynamics only if the unitarity breaking term is both stochastic and non-linear. In the main text, we therefore consider unitarity breaking perturbations of the form of Eq.~\eqref{Eq:key}. These influence the dynamics of superposed states like that of Eq.~\eqref{eq:finalvN}, resulting from the entanglement of a microscopic system with the pointer of a macroscopic measurement apparatus. Pointers (of any sort) are necessarily symmetry-broken objects, and the states resulting from spontaneous symmetry breaking are necessarily pointer states in the sense of being stable against environmental decoherence~\cite{Zurek_1981}. 

The time evolution of superposed pointer states has a \emph{diverging susceptibility} to non-unitary perturbations in the thermodynamic limit, as signalled by Eq.~\eqref{eq:XtalSUV}. The result is a near-instantaneous evolution towards a single pointer state, indicating a single measurement outcome. As in the standard theory for spontaneous symmetry breaking, the collapse does not arise from nothing, but the presence of a mathematical divergence in the thermodynamic limit indicates that for realistic, physical sizes of measurement machines, exceedingly small non-unitary perturbations suffice to cause collapse dynamics that is for all practical purposes unpredictable, inevitable, and instantaneous.

Notice that we do not make predictions in the current work about the precise time evolution to be expected in any particular mesoscopic experiment. Also, we do not estimate any values for the model parameters we use. Rather, we show that spontaneous unitarity violation can give rise to the emergence of Born’s rule, and that it emerges as a collective effect in the dynamics of very large systems exposed to a very weak non-unitary perturbation. The emergence is spontaneous in the sense that it is unavoidable, yet unpredictable, due to a formally diverging susceptibility in the limit of large system size. It is universal in the sense that Born’s rule will arise for sufficiently large systems from their extremely weak interaction with a non-unitary stochastic field, regardless of the precise interaction strength. Finally, Born’s rule emerges rather than being imposed or assumed, in the sense that the stochastic fluctuations leading to it are taken from a flat distribution, without any knowledge of the state being measured.

\section{Strong measurement \label{app:vonNeumann}}
In this appendix, we motivate the use of Eq.~\eqref{eq:init_superposition} in Sec~\ref{sec:0} by summarizing the strong measurement setup originally introduced by Von Neumann~\cite{Von_Neumann2018-bo}. We consider a measurement apparatus $A$, performing a single measurement of an observable $\hat{O}_S$ on the system $S$. We assume quantum theory to apply to the measurement device as well as the system, and consider their combined quantum mechanical time evolution. 

For concreteness, we consider a measurement device with a  pointer whose centre-of-mass position $x$ along a dial will indicate the measurement outcome. This does not lead to any loss of generality, as the pointer can be replaced with any type of classical state arising from a spontaneously broken symmetry (see appendix~\ref{app:SSB}). In that case, $x$ should be considered an eigenvalue of the order parameter operator~\cite{Wezel_SSBlecturenotes}, and the corresponding eigenstates will be classical symmetry-broken states that are guaranteed to be stable under environmental decoherence and can thus be considered `pointer states' in the general sense~\cite{Zurek_1981}. Notice that the use of pointer states with a spontaneously broken symmetry is necessitated by the fact that only these states are susceptible to spontaneous unitarity violation~\cite{vanwezelprb}. This introduces a preferred basis for measurement outcomes, which must always be eigenstates of an order parameter operator.
 
The state of the measurement apparatus, $\ket{\psi}_A$, may be expressed in a basis of states $\ket{x}_A$ with fully localised centres of mass $x$ for the pointer:
\begin{align}
\ket{\psi}_A = \int dx ~ \psi(x) \,\ket{x}_A
\end{align}
Considering the pointer to be a macroscopic object in a symmetry-broken coherent state~\cite{Wezel_SSBlecturenotes}, the initial state of the pointer wave function $\psi(x)$ before measurement will be given by a sharply peaked Gaussian of the form:
\begin{align}
\psi(x) = \left(\frac{1}{2\pi \Delta^2}\right)^{\frac{1}{4}}e^{-x^2/4\Delta^2}.
\label{Eq:pointer}
\end{align}
For a measurement apparatus that is sufficiently large to spontaneously break a symmetry, the spread $\Delta $ will be exceedingly small~\cite{Wezel_SSBlecturenotes}. Wave functions centered at different, well-separated positions then have an exponentially small overlap, and can be used to unambiguously resolve different measurement outcomes. For simplicity, we consider the system observable $\hat{O}_S$ to have a discrete spectrum of eigenstates $\ket{\sigma}_S$ with eigenvalues $\sigma$. 

For the apparatus to function as a measurement device, the Hamiltonian governing the interaction between system and apparatus should be such that initial system states $\ket{\sigma}_S$ with different values of $\sigma$ cause the pointer to evolve to different centre of mass positions. This is accomplished by a generic interaction Hamiltonian of the form:
\begin{align}
\hat{H}_{\text{int}} = \gamma \hat{O}_S \otimes \hat{P}_A.
\end{align}
Here, $\gamma$ is the strength of the interaction and $\hat{P}_A$ is the canonical momentum operator conjugate to the pointer position, so that $[\hat{X}_A,\hat{P}_A]=i\hbar$. The time evolution operator generated by this Hamiltonian acts as a shift operator on the pointer position, with the size of the shift determined by the eigenvalue of the system observable:
\begin{align}
e^{-\frac{i}{\hbar} t \hat{H}_{\text{int}}} \ket{\sigma}_S \ket{\psi}_A = \int dx ~ \psi(x - \sigma {\gamma t }/{\hbar}) \,\ket{\sigma}_S \ket{x}_A
\end{align}

In a generic measurement process, the system will be in a superposition of multiple eigenstates of $\hat{O}_S$ before measurement. The combined initial state of system and apparatus is then of the form $\ket{\Psi(t=0)}_{SA} = \sum_\sigma \phi_\sigma \ket{\sigma}_S \ket{\psi}_A$. Unitarily evolving with the time evolution generated by the interaction Hamiltonian then causes the formation of macroscopic entanglement: 
\begin{align}
    \ket{\Psi(t)}_{SA}= \sum_\sigma \int dx ~  \phi_\sigma \, \psi(x - \sigma {\gamma t }/{\hbar})  \,\ket{\sigma}_S \ket{x}_A
    \label{Eq:VNS_endstate}
\end{align}
In more quantitative modeling one may consider a realistic time dependent impulse function $\gamma(t)$ instead of the constant $\gamma$ used here, but this only affects the speed with which the evolution unfolds and not the qualitative formation of entanglement between system and apparatus. Notice that each of the states $\ket{\sigma}_S$ becomes entangled with its own pointer state sharply peaked around the spatial position $x_\sigma(t) = \sigma {\gamma t }/{\hbar}$. The qualitative formation of entanglement is instantaneous, but the amount of entanglement grows with time as the pointer states centered at different $x_\sigma(t)$ separate from one another.

After some time, the final state  obtained in Eq.~\eqref{Eq:VNS_endstate} is of the same form as Eq.~\eqref{eq:init_superposition} in Sec.~\ref{sec:0} of the main text: 
\begin{align}
\ket{\Psi}_{SA} &=\sum_\sigma \phi_\sigma \ket{\sigma}_S\ket{x_\sigma}_A, \notag \\
\text{with} ~~~ \ket{x_\sigma}_A &=\int dx\, \psi(x - x_\sigma) \,|x\rangle_A. 
\label{eq:finalvN}
\end{align}

According to Born's rule, $|\phi_\sigma|^2$ gives the probability of obtaining any one of the classical pointer states $\ket{x_\sigma}_A$ upon performing the measurement. The models for spontaneous unitarity violation considered in the main text start from the initial state of Eq.~\eqref{eq:finalvN} and explain its probabilistic reduction to just one component $\ket{\sigma}_S\ket{x_\sigma}_A$. 

It would be possible to formally separate the unitary entangling dynamics from the non-unitary quantum state reduction if either the strength of the non-unitary perturbation does not depend on the amount of overlap between distinct pointer states, or if the entanglement dynamics is completed instantaneously. Neither is a realistic assumption for real measurements. However, since the evolving overlap will only affect the speed at which the non-unitary time evolution unfolds and not its final state, the assumption of instantaneous separation between pointer states does not influence the statistics of measurement outcomes that are the focus of the current work.

\bibliography{biblio}


\end{document}